\documentclass[10pt,aps,prd,floatfix,nofootinbib,preprintnumbers,groupedaddress]{revtex4-2}
\usepackage{comment}
\usepackage{graphicx}
\usepackage{amsmath, mathrsfs}
 \usepackage{xcolor}
 \usepackage{amssymb}
\usepackage{bbm}
\usepackage{csquotes}
\usepackage{tensor}

\usepackage{tikz}
\usetikzlibrary{positioning,decorations.pathmorphing}

\usepackage{epstopdf}
\newcommand{\bea}{\begin{aligned}}
\newcommand{\eea}{\end{aligned}}
\newcommand{\beq}{\begin{equation}}
\newcommand{\eeq}{\end{equation}}

\newcommand{\bse}{\begin{subequations}}
\newcommand{\ese}{\end{subequations}}

\makeatother
\def\beq{\begin{eqnarray}}
\def\eeq{\end{eqnarray}}
\def\be{\begin{equation}}
\def\eeq{\end{equation}}
\def\={\triangleq}
\def\nn{\nonumber\\}

\def\mb#1{\mathbb{#1}}
\def\mbf#1{\mbox{\boldmath${#1}$}}

\def\lie{\pounds}
\def\l{\ell}

\usepackage{hyperref}
\hypersetup{
     colorlinks   = true,
     citecolor    = red,
     linkcolor    = blue,
     urlcolor     = blue,
}

\newcommand{\bmm}{\begin{multline}}
\newcommand{\emm}{\end{multline}}

\def\={\stackrel{\Delta}{=}}
\def\ula#1{\underleftarrow{#1}}
\def\lie{\pounds}
\def\l{\ell}

\begin{document}
\title{Non-Minimally Coupled Scalar Field, Area Quantization and Black Hole Entropy}

\author{Akriti Garg}
\email{akritigarg571@gmail.com}
\affiliation{Department of Physics \& Astronomical Science, Central University of Himachal Pradesh, Dharamshala-
176215, India.}

\author{Sahil Devdutt}
\email{devduttsahil@gmail.com }
\affiliation{Department of Physics \& Astronomical Science, Central University of Himachal Pradesh, Dharamshala-
176215, India.}

\author{Ayan Chatterjee}
\email{ayan.theory@gmail.com}
\affiliation{Department of Physics \& Astronomical Science, Central University of Himachal Pradesh, Dharamshala-
176215, India.}

\pagenumbering{arabic}
\renewcommand{\thesection}{\arabic{section}}
\begin{abstract}
    The enumeration of black hole entropy in candidate theories of quantum gravity 
    utilises the quantum properties of microstates residing on the black hole horizon. For example, in Loop Quantum Gravity, the computation of entropy is based on the spectrum of area operator, and one determines the possible number of area microstates corresponding to a given classical horizon area. In this paper, we derive the eigenspectrum of the horizon area operator for rotating/non-rotating black holes in a gravitational theory non-minimally coupled to scalar fields. Using the weak isolated horizon formalism, we show that the spectrum of area operator follows unambiguously from 
    the algebra of horizon symmetry. 
    More precisely, from the quantum mechanical point of view, the horizon geometry must be naturally discrete, a conclusion which is arrived at directly, without the need for any particular theory of quantum gravity. The area spectrum depends on the Barbero-Immirzi parameter as well as the value of scalar field on horizon. The area spectrum is equidistant, which is consistent with the Bekenstein-Mukhanov proposal and gives rise to black hole entropy and their quantum corrections.
\end{abstract}

\maketitle

Understanding the classical and quantum properties of black hole horizon, and black hole entropy in particular, remains one of the most challenging problems of gravitational physics \cite{Wald:1995yp, Wald:1984rg, Bekenstein:1994bc, Sorkin:2005qx}. It is strongly believed that a better knowledge of 
the quantum properties of horizon may allow us a peek into the microscopic world of quantum spacetime \cite{Wheeler1992ItFromBit, Bekenstein:1995ju, Mukhanov:2018tzo, Perez:2023ctt}. Furthermore, one also expects that the quantum formulation of geometry/gravity may eventually help to resolve the problems associated with classical gravitational physics \cite{Ashtekar:2004eh, Perez:2017cmj, Bojowald:2007ky, Hawking1996THENO, Penrose2004RoadToReality, Kiefer:2004xyv, Frolov:1998wf}. Therefore, the first test of any candidate theory of quantum gravity is whether it successfully determines 
the value of black hole entropy, consistent with 
the first law of black hole mechanics. For general theory of relativity (GR), the black hole entropy is proportional to the area of horizon cross-section \cite{Bardeen:1973gs, Bekenstein:1973ur, Hawking:1971tu}.  This value of entropy, together with Hawking's derivation of black hole radiance, provides consistency to Bekenstein's claim that the laws of black hole mechanics must be same as the laws of thermodynamics \cite{Hawking:1975vcx, Hawking:1974sw}. More precisely, if quantum mechanics is taken into account, a black hole in GR with surface gravity $\kappa$, must be considered as a thermal object with temperature $\mathcal{T}=(\kappa/2\pi)$, and entropy $\mathcal{S}=\mathcal{A}/4\ell_{p}^{2}$, where 
$\mathcal{A}$ is the area of horizon cross-section
and $\ell_{P}$ is the Planck length. Note that with these identifications, the zeroth law, which states that the surface gravity must be constant on the horizon, becomes identical to the zeroth law of thermodynamics whereas, the first law of horizon mechanics, for a black hole of mass $\mathcal{M}$, is mapped to the first law of thermodynamics $d\mathcal{M}=\mathcal{T}d\mathcal{S}$. The second law of black hole mechanics, which states that the
horizon area of a classical black hole must 
always increase, provides credence to the claim that $\mathcal{S}$ is proportional 
to $\mathcal{A}$ \cite{Hawking:1971vc,Sarkar:2019xfd}. The two leading theories of quantum gravity, String Theory and Loop Quantum Gravity (LQG) have successfully obtained this leading value of entropy from the enumeration of horizon microstates \cite{Dabholkar:2004yr, Dabholkar:2005dt, Dijkgraaf:1996it, Dabholkar:2008zy, Mandal:2010cj, Ashtekar:1997yu, Kaul:2000kf, Meissner:2004ju, Domagala:2004jt, Smolin:1995vq, Ashtekar:2000eq, Rovelli:1996dv, Krasnov:1996tb, Ghosh:2004wq}. For string theory, the microstate computations have been carried out for certain $4$-dimensional supersymmetric ($\mathcal{N}=4$ and $\mathcal{N}=8$) BPS black holes and five- dimensional BMPV solutions, whereas for LQG, the computations are for extremal or 
non- extremal (weak isolated) horizons in $4$- dimensions \cite{Mandal:2010cj,Ashtekar:2004cn}. Furthermore, subleading corrections to entropy have also been obtained from such counting techniques and have provided fundamental clues towards quantum properties of black hole horizon \cite{Ghosh:2004rq, Kaul:2000kf,Meissner:2004ju,Domagala:2004jt,Ghosh:2004wq, Ghosh:2011fc, Ghosh:2013iwa, Perez:2023ctt}. Both of these quantum theories agree that the general expression of black hole entropy, obtained from the microscopic computation, should have the following expansion (for large black holes, $\mathcal{A} \gg \ell_{P}^{2}$):\\
\begin{equation}\label{entroy_expression}
    \mathcal{S}= \frac{\mathcal{A}}{4\ell_{P}^{2}}+ a_{1}\, \ln \frac{\mathcal{A}}{\ell_{P}^{2}} +
    a_{2}\,  \frac{\ell_{P}^{2}}{\mathcal{A}} +\cdots ,
\end{equation}
where $a_{i}$ are coefficients in the expansion, but there is no general consensus on their values \cite{Ghosh:2004rq,Kaul:2000kf,Meissner:2004ju,Domagala:2004jt,Sen:2012dw,Ghosh_2005, PhysRevLett.102.141302}. This general expectation for the value of entropy, eqn. \eqref{entroy_expression}, also follows from macroscopic computations carried out through the Wald formula or the the entropy function formalism \cite{Wald:1993nt,Iyer:1994ys,Jacobson:1993vj,Sen:2005wa,Sen:2007qy}. For small black holes, exponential corrections to the area value also arise for the black hole entropy \cite{Chatterjee:2020iuf,Dabholkar:2014ema}. Similar results for the leading and subleading corrections to black hole entropy have also been obtained from the AdS- CFT perspective with asymptotic and near horizon symmetries, Euclidean quantum gravity, and other semiclassical techniques like entanglement, the brick wall approach, or the tunneling methods \cite{Strominger:1996sh,Strominger:1997eq,Brown:1986nw,Brown:1995su,
Carlip:1998wz,Carlip:1999cy,Carlip:2000nv,Carlip:2002be,Dreyer:2001py,
Gupta:2001bg,Mukherji:2002de,Medved:2004eh,Page:2004xp,Banerjee:2008cf,
Banerjee:2008fz,Majhi:2008gi,Cai:2009ua,Aros:2010jb,Gibbons:1976ue,
Solodukhin:1994st,Solodukhin:1994yz,Fursaev:1994te,Mavromatos:1995kc,
Mann:1996bi,Mann:1997hm,Govindarajan:2001ee,Solodukhin:2010pk,
Banerjee:2010qc,Banerjee:2011jp,Bombelli:1986rw,Sarkar:2008zzb,tHooft:1984kcu,Mitra:2009ba,Hayward:2008jq}.   \\

The laws of black hole mechanics change if a 
scalar field ($\phi$) is present in 
the spacetime, which couples not just to 
the spacetime metric, but also to 
the Ricci scalar through a function $f(\phi)$. Naturally, the classical first law for black holes in 
this theory is altered, and the 
horizon area is modified so that 
the entropy must be identified with $f(\phi_{\Delta})\mathcal{A}$, where $\phi_{\Delta}$ being value of the field on
the horizon \cite{Ashtekar:1999sn,Chatterjee:2009vd,Ashtekar:2003jh}, provided 
the temperature is $\kappa/2\pi$. The 
question now is to provide a microscopic derivation 
of this entropy value through counting of
horizon microstates. The theory of LQG does provide some clues towards this value \cite{Ashtekar:2003zx}. Since this method 
shall play an important role for the purposes of 
comparison with our approach, in the following, we provide a brief discussion 
of the LQG computation. \\

The calculation proceeds through several steps \cite{Ashtekar:1999sn,Chatterjee:2009vd,Ashtekar:2003jh,Ashtekar:2003zx, BarberoG:2015xcq}: First, using the Weakly Isolated Horizon (WIH) formalism, one constructs the classical phase space of black hole solutions of the gravitational theory with a non- minimally coupled scalar field (a discussion on 
the notion of WIH shall be given in the following sections). This space of solution contains all those black hole spacetimes whose inner boundary (which we shall denote by $\Delta$) satisfies the conditions of a WIH and are asymptotically flat at spatial infinity (the asymptotic conditions do not play particular role in the entropy computation and we shall not devote any further discussions on this point). More precisely, in four dimensions, $\Delta$ is assumed to be an expansion- free null hypersurface with topology $\mathbb{S}^{2}\times \mathbb{R}$ on which the equations of motion and the (dominant) energy conditions 
hold true.
One may show that the null vector $\ell^{a}$ generating $\Delta$ is a Killing vector field \emph{on} the horizon; although the immediate neighbourhood may be non- stationary, which may not even admit any timelike Killing vector. If one restricts to spherically symmetric sector of 
the phase- space, the gravitational symplectic structure cleanly separates into two parts, (a) a bulk part consisting of the conjugate pair of tetrads and spin- connections, and (b) a boundary term which is completely determined through a one-form $V$.
For non- rotating and non- extremal black holes, 
this boundary theory turns out to be 
topological, and more precisely, a $U(1)/SU(2)$ Chern- Simons theory corresponding to this one- form $V_{a}$ \cite{Engle:2009vc, Engle:2010kt, Kaul:2010kg}. The level of 
the Chern- Simons theory depends on 
the value of scalar field on horizon, and is given by $[f(\phi_{\Delta})\mathcal{A}/4\pi\gamma\ell_{P}^{2}]$, where $\gamma$ is the Barbero- Immirzi parameter. This completes the first step, where one associates, with the horizon $\Delta$, a field theory whose quantisation may provide the states responsible for obtaining the entropy. Note that
matter fields, like the Maxwell or Yang- Mills, or minimally coupled scalar fields, do not contribute to the surface symplectic structure, the Chern- Simons field $V_{a}$ is purely gravitational. \\

The second step is to quantise this spherically -symmetric sector of the phase- space \cite{Ashtekar:2003zx,Ashtekar:2004eh,Ashtekar:1997yu,Ashtekar:2000eq}. One assumes that 
the (kinematical) Hilbert space is a tensor product of the bulk ($\mathcal{H}_{V}$) and the surface ($\mathcal{H}_{S}$) Hilbert spaces with the form $\mathcal{H}=\mathcal{H}_{V}\otimes \, \mathcal{H}_{S}$ (Note that this is an assumption in line with the fact that the classical symplectic structure also has a similar decomposition). $\mathcal{H}_{V}$ is  Hilbert space of states of the bulk and describe polymer excitations of geometrical and matter fields. More precisely, $\mathcal{H}_{V}$ is spanned by \emph{spin networks}, denoted by $|\mathscr{G}; j, m \rangle$, which are characterised by a graph $\mathscr{G}$, with $j$ labelling the edges and $m$ labelling the vertices. The spin network is a particularly convenient basis since they are eigenstates of the area operator. In particular, given a closed surface
$\mathbb{S}$ in the bulk, one may view the classical area of this surface as arising from the quantum excitations generated due to intersection of the edges of $\mathscr{G}$ with $\mathbb{S}$. So, one may view edges as depositing quantised areas when they puncture $\mathbb{S}$. If there are $n$ edges which puncture $\mathbb{S}$, the eigenvalue of the area operator is $[8\pi\gamma\ell_{P}^{2}/f(\phi_{\Delta})]\,\sum_{i=1}^{n}\sqrt{j_{i}(j_{i}+1)}$ \cite{Ashtekar:1997yu}. On the other hand, surface Hilbert space $\mathcal{H}_{S}$ contains space of states of the Chern- Simons theory on a punctured sphere, along with the prequantisation condition that $[f(\phi_{\Delta})\mathcal{A}/4\pi\gamma\ell_{P}^{2}]$ must
be integer valued. To obtain the physical Hilbert space,
one must reproduce the classical embedding of 
the horizon $\Delta$ in the bulk as a quantum condition. As a result, the states belonging to the physical Hilbert space
must be such that they satisfy 
 constraints in the bulk and 
the \emph{quantum horizon condition} (sometimes called 
 $F-\Sigma$ equation) on the horizon. Let us summarise 
the picture of a quantum horizon $\Delta_{Q}$: The states which characterise $\Delta_{Q}$ are the spin network states. Let a graph $\mathscr{G}$ of this spin- network with edge labelled by $j_{i}$, punctures the horizon cross- section $S_{\Delta_{Q}}$, with label $m_{i}$ [with the requirement that $m_{i}$ has $(2j_{i}+1)$ values, $-j_{i}\le m_{i}\le j_{i}$], then area of the quantum horizon $\mathcal{A}_{Q}$ is taken to be the value of area which the bulk area operator assigns to punctures on $S_{\Delta_{Q}}$, so that $\mathcal{A}_{Q}=[8\pi\gamma\ell_{P}^{2}/f(\phi_{\Delta})]\,\sum_{i=1}^{n}\sqrt{j_{i}(j_{i}+1)}$, such that
the additional \emph{total projection} constraint, 
 $\sum_{i} m_{i}=0$ is also satisfied. This completes 
the characterisation of the quantum horizon, and we are ready for the third step, that of state counting.
The third and final step is to obtain the entropy using 
the point of view of microcanonical ensemble. Given the classical area $\mathcal{A}$, the counting must obtain the number of states $\mathcal{N}$ such that for a small value of $\delta$:
\begin{eqnarray}\label{area_lqg}
    \mathcal{A}-\delta\le [8\pi\gamma\ell_{P}^{2}/f(\phi_{\Delta})]\,\sum_{i=1}^{\mathcal{N}}\sqrt{j_{i}(j_{i}+1)}\le \mathcal{A}+\delta .
\end{eqnarray}
Assuming the black hole area is large, the entropy is enumerated from $\mathcal{S}=\ln \mathcal{N}$, (using $k_{B}=1$ units) 
through which one obtains the
equation eqn. \eqref{entroy_expression} with certain choice of parameter $\gamma$ and $a_{1}=-1/2$.  \\

How does this calculation fare in comparison to other methods?
Do the leading and subleading values of entropy match with the LQG method\footnote{One must point out that even within the LQG framework, there are competing claims on the coefficient of the log term, depending on whether one uses the $U(1)$ or the $SU(2)$ Chern- Simons theory or how different spin configurations are being counted \cite{Kaul:2000kf,Meissner:2004ju,Ghosh:2004wq,Ghosh_2005,Ghosh:2004rq,Domagala:2004jt}.} ? The answer is both yes and no, which is natural since the methodologies for computing black hole entropy are vastly different. Over the years, several techniques have been developed to compute black hole entropy,
and to understand its precise dependence of the horizon area.
The first attempt has been through the Euclidean quantum gravity approach \cite{Gibbons:1976ue}. This method has
also been used to obtain quantum corrections to black hole entropy including the logarithmic terms, with coefficients matching the LQG results \cite{Solodukhin:1994st,Solodukhin:1994yz,Fursaev:1994te,Mavromatos:1995kc,Mann:1996bi,Mann:1997hm,Govindarajan:2001ee, Rovelli:2011eq}, although recent careful studies have disputed this claim \cite{Banerjee:2010qc,Banerjee:2011jp,Sen:2012kpz,Gupta:2014hxa,Keeler:2014bra,Larsen:2014bqa,Karan:2019gyn,Sen:2012cj,Bhattacharyya:2012wz,Sen:2012dw}. Secondly, a popular approach to calculate black hole entropy from first principles rely on conformal field theory techniques, and the use of Cardy formula. The CFT on the boundaries (both at asymptopia and at the horizon) are obtained from the algebra of vector fields which preserve the corresponding asymptotic symmetries \cite{Strominger:1996sh,Strominger:1997eq,Brown:1986nw,Carlip:1999cy,Carlip:1998wz,Carlip:2002be,Dreyer:2001py}. Here also, the leading and the logarithmic corrections to the Cardy formula provide a successful matching with the LQG value of black hole entropy. On the holographic approach, the Ryu- Takayanagi method attempts to obtain the black hole entropy through geometry and computes entanglement entropy in the boundary CFT through bulk minimal surfaces, where the logarithmic terms appear due to high-order metric or minimal surface deformations in the bulk \cite{Ryu:2006bv}. The semiclassical techniques of brick- wall approach is another popular method which has provided a successful matching with the leading and subleading terms in LQG \cite{Sarkar:2008zzb,tHooft:1984kcu}, as do others where the logarithmic corrections are related to the trace anomaly \cite{Banerjee:2008fz,Cai:2009ua}. However, the most popular approach to understand \emph{macroscopic} black hole entropy has been the  Wald formula \cite{Wald:1993nt,Iyer:1994ys,Jacobson:1993vj,Aros:2010jb}, and the entropy function formalism \cite{Sen:2007qy}. For these approaches too, computations have shown a general trend of eqn. \eqref{entroy_expression}, although coefficients are often widely different from those of LQG. \\

Let us now come back to our discussion of the LQG method and if the approach may be improved upon. Note that, given the developments in the LQG literature, there exists some major gaps which we wish to address here so as to provide a better understanding of the origin of black hole entropy.
Let us enumerate the list of issues. First, the proof that the surface part of the gravitational symplectic structure (defined on the classical phase space admitting a WIH) is that of a $U(1)/SU(2)$ Chern- Simons theory, requires that the horizon cross- section be 
spherically symmetric. Quite remarkably, distorted and horizons with multipole structures have been studied as well \cite{Ashtekar:2004nd,Perez:2017cmj}, although the computations become quite involved, and a simpler method may provide better insights.  Secondly, during the state counting, when the classical area is matched with 
the eigenvalue of area operator [given by $8\pi\gamma\ell_{P}^{2}\sqrt{j(j+1)}$ ], one uses the expression obtained from the bulk. More precisely, this expression of area eigenvalue is evaluated for a surface $\mathbb{S}$ which foliate the spacelike hypersurface $\Sigma$ on the full spacetime. Therefore, during the LQG calculation, one has to assume that eigenvalue of the area operator shall continue to hold true for a sphere foliating the null hypersurface as well. Naturally, a direct evaluation of 
the spectrum of the area operator on horizon should give us a better perspective into the nature of horizon microstates. Thirdly, the evaluation of horizon entropy requires that the black hole has large area and indeed, the leading and subleading values are evaluated using the Stirling approximation with $\mathcal{A}\gg \ell_{P}^{2}$. Therefore, one needs to understand the changes one may encounter if this \emph{large area constraint} is removed. Studies on such Planck area sized black holes have revealed important features like oscillations in the entropy spectra with constant periodicity \cite{Corichi:2006wn}, which points to an effective model of equidistant spectrum for entropy as envisaged through the Bekenstein- Mukhanov model \cite{Bekenstein:1995ju}.  Fourth, and the one which we believe is an important question, is the following: Is it possible to develop a microscopic model of black hole horizon, without using any formulation of quantum gravity (either LQG or String Theory) through which the origin of black hole entropy be explained? How much is the area operator and its spectrum tied to the LQG formulation? Or in other words, is it possible to obtain a quantised area operator on the horizon independent of the theory of quantum gravity?  We shall show that it is indeed possible to develop a model based on symmetries of the horizon where all these questions may be answered in the affirmative and 
the issues be resolved quite naturally.\\

We shall use the method advocated in \cite{Chatterjee:2020iuf} for GR, which we briefly review. First, one constructs the classical phase space of the first order Holst action admitting WIH as an inner boundary (and is asymptotically flat at spatial infinity). The second step is to note that the WIH boundary, which admits a local Lorentz symmetry of $\text{ISO}(2)\ltimes \mathbb{R}$, gives rise to observable charges proportional to the area of horizon cross- section. This is not unexpected since it is well known that local symmetries in gauge theory give rise to observable edge states on boundaries \cite{Witten:1988hf,Balachandran:1991dw,Ashtekar:1999wa}.
On the WIH, the observable (area of horizon cross- section, $\mathcal{A}$) is Hamiltonian charge due to the rotation and boost generators corresponding to local $\text{ISO}(2)\ltimes \mathbb{R}$ transformations. Thirdly, one may note that the algebra of $\text{ISO}(2)\ltimes \mathbb{R}$ dictates that the angular momentum generators must be quantised with integers  $n\hbar$, and quantum states residing on the horizon cross-sections carry a 
representation of $\text{ISO}(2)$.
Since the horizon area $\mathcal{A}$ is also the generator of rotations, this provides a link to label the quantum states of the WIH by the integers $n\hbar$ or, in the language of quantum states, the eigenstates of the area operator. More precisely, one may visualise the classical horizon being formed out of quantised areas. The fourth step is to use the notion of microcanonical ensemble and determine the number of 
ways one may assimilate the quantised areas to obtain a value of classical area, the logarithm of which should provide the measure of entropy. Naturally, this counting also leads to the correct value of entropy and area but
does not provide log corrections \cite{Ghosh:2012jf}. Our approach shall be to 
use these techniques to evaluate the black hole entropy, although major changes shall arise due to the presence of the non- minimal scalar field. In particular, we shall show that
the area spectrum does depend on this scalar field and hence, the quantisation and entropy counting does explicitly carry the imprint of this scalar field. As a result, the entropy shall depend both on the horizon area and the value of scalar field on $\Delta$.\\

We proceed as follows: In the next section, we discuss the boundary condition of a WIH, and develop the zeroth and the first laws of black hole mechanics for $N$- dimensions using the first order non- minimally coupled Palatini action based on tetrads and spin- connections (although we will be working in $N$-dimensions, we shall continue to call the orthogonal basis as tetrads). 
This action describes a gravitational theory coupled non- minimally to a scalar field.
We shall construct the phase -space of this theory, and obtain the symplectic structure on the space of solutions
which admits a WIH as the internal boundary. We shall show that the zeroth law of horizon mechanics arises a 
result of the boundary conditions. We shall also derive the first law of horizon mechanics and show that the necessary and sufficient condition for the existence of 
first law of horizon mechanics is that a well defined Hamiltonian evolution is generated on the phase space. Moreover, the first law shall also show that, if thermodynamics is to be taken into account, one must include the contribution of scalar field towards the black hole entropy. Then, we discuss the Holst action for the $4$- dimensional theory, where an additional term must
be added to the Palatini action. This additional Holst term does not contribute to the first law of horizon mechanics but does alter the symplectic structure when boundaries are present. Using this new Holst symplectic structure, we shall derive the Hamiltonian charges corresponding to the local Lorentz rotation and boost transformations on the horizon $\Delta$, and show that they are proportional to the cross- sectional area of $\Delta$. The next section contains a discussion on the quantum theory, the area spectrum and the calculation of black hole entropy. A discussion is carried out in the last section.



\section{Horizon geometry and mechanics in 
\texorpdfstring{$N$-dimensions}{N-dimensions}}
We begin by providing a motivation for developing 
the WIH formalism (a detailed discussion may be found in \cite{Ashtekar:1998sp,Ashtekar:2004cn,Ashtekar:2000sz,Ashtekar:1999yj,Ashtekar:2000hw,Ashtekar:2025wnu}). The power of the WIH formalism lies in the fact that it provides a realistic framework for evolution of black hole horizons, along with a broader foundation to understand the zeroth and the first laws of black hole mechanics. 
The WIH is a quasilocal notion of a stationary black hole horizon which, unlike the event horizon formalism, does not require knowledge of the future null asymptotic infinity to ascertain its location. A WIH is capable of describing a stationary horizon in a non- stationary spacetime, provided no energy- momentum crosses this surface. As a result, 
the space of solutions which admit a WIH is much larger compared to that of an event horizon.  The WIH may
be modeled as a  codimension-$1$ null tube foliated by marginally trapped surfaces, and in the following, we shall develop this formalism in a similar fashion. More precisely, we shall consider a generic null hypersurface ($\Delta$) in the spacetime, and impose certain conditions on $\Delta$ such that it acquires the properties of a stationary horizon.\\

Let us consider a $N$- dimensional spacetime $\mathscr{M}$ endowed with metric $g_{ab}$ and signature $(-++\cdots)$, along with a metric compatible covariant derivative $\nabla_{a}$. We shall use 
the orthonormal basis for our calculations and hence develop the Newman- Penrose 
formalism for higher dimensions, which may be easily adapted to the null hypersurfaces. 
We use the following notation 
for vectors and one-forms respectively \cite{Ortaggio:2007eg,Chatterjee:2025jur}: 
\begin{align}\label{ortho1}
    m^{(0)}&=n & m^{(1)}&=\ell & m^{(i)}&= \text{spacelike covectors}\\ \label{ortho2}
    m_{(0)}&=-\ell & m_{(1)}&=-n &  m_{(i)}&= \text{spacelike vectors},
\end{align}
such that the following orthonormality conditions hold: $\ell\cdot \ell=0,\, n\cdot n=0\, , \ell\cdot n=-1,\, \ell\cdot m^{(i)}=0,\, n\cdot m^{(i)}=0,\, m^{(i)}\cdot m^{(j)}=\delta^{(ij)}$. Note 
that a common label $m^{(a)}$, such that 
the indices $a,b,c$ run from $0$ to $(D-1)$, is used for all 
the frame vectors. The indices $i,j,k$ run from $2$ to 
$(N-1)$, and therefore represent only spacelike vectors $m^{(i)}$. The spacetime metric is given by:
\begin{equation}
    g_{ab}=-2\ell_{(a}n_{b)}+m_a^{(i)} m_{b}^{(j)} \delta_{(ij)}.
\end{equation}
Let $\Delta$ be a $(N-1)$ dimensional null hypersurface in $\mathscr{M}$, generated by a future directed null normal $\ell^{a}$, and endowed with a degenerate metric $q_{ab}\=g_{\ula{ab}}$ (the symbol $\=$ shall denote equalities which hold only on $\Delta$, and the underarrows show the indices are pulled back to $\Delta$). We assume that $\Delta$ is foliated 
by spacelike $v=$constant 
$(N-2)$-surfaces $S_{v}$. 
For a simplified Newman- Penrose like description in higher dimensions, we 
choose the two null normals to $S_{v}$ 
as $\ell^{a}$ and $n^{a}$, with the one-form $n_{a}$ being such that $n_{a}=-(dv)_{a}$. The vector field  $\ell^{a}=(\partial_{v})^{a}$ is then tangent to $\Delta$. The 
spacelike vectors $m_{(i)}^a$ form the basis spacelike
cross-sections of $S_{v}$. Let us define the following Ricci rotation coefficients:
\begin{eqnarray}\label{Ricci_coeff}
    \nabla_{b} \,\ell_{a} &=& L_{cd}\, m_a^{(c)}m_b^{(d)}\, ,\label{100}
\end{eqnarray}
These Ricci coefficients are 
not all independent or non- zero. To obtain 
the constraints on these functions we first use 
the orthonormality condition $\ell\cdot\ell=0$ 
to obtain:
\begin{eqnarray}\label{Ricci_cond}
   L_{0a} = 0, 
\end{eqnarray}
Second, since $\Delta$ is null, 
the generator $\ell^{a}$ must be geodetic. 
From equation eqn.\eqref{Ricci_coeff}, 
\begin{equation}\label{Ricci_cond_1}
    \ell^b\nabla_b \ell_{a}=L_{c0} m_a^{(c)}=L_{10}\ell^{a}+L_{i0}m^{(i)}_{a}.
\end{equation}
Since the geodetic condition requires that 
the right side be proportional to $\ell^{a}$, it then follows 
that $L_{i0}\stackrel{\Delta}{=}0$.
Using $L_{0a}=0$ and $L_{i0}=0$ we can now write the eqn. \eqref{Ricci_coeff} as:
\begin{equation}\label{gradleqn}
    \nabla_{\underleftarrow{b}}\ell^{a} \stackrel{\Delta}{=}\left[L_{10} n_b + L_{1i} m_b^{(i)}\right] \,\ell^{a} + L_{ij}m_b^{(j)}m^{(i)\,a}.
\end{equation}
    The matrix $L_{ij}$ can be decomposed in the symmetric (and tracefree) and the antisymmetric parts in the following manner: $L_{ij}=S_{ij}+A_{ij}$, where $S_{ij}=L_{(ij)}=\sigma_{ij}+\theta/(N-2)\delta_{ij}$ and $A_{ij}=L_{[ij]}=\omega_{ij}$ is the twist corresponding to the generator $\ell^{a}$. If 
    the surface $\Delta$ is twist-free, shear- free and expansion- free null surface, 
    then $L_{ij}\stackrel{\Delta}{=}0$, and it further follows from eqns. \eqref{Ricci_cond}, and \eqref{Ricci_cond_1}, that the following quantities vanish identically on $\Delta$:\\
\begin{equation}
    L_{0a}\stackrel{\Delta}{=}L_{i0}\stackrel{\Delta}{=}L_{ij}
    \stackrel{\Delta}{=}0. 
\end{equation}
So, once the dust settles, the eqn. \eqref{gradleqn} may be written as:
\begin{equation}\label{gradleqn1}
    \nabla_{\underleftarrow{b}}\ell^{a} \stackrel{\Delta}{=}\omega_{b}\, \,\ell^{a},
\end{equation}
with $\omega=\left[L_{10} n + L_{1i} m^{(i)}\right]$.
This implies that $\underleftarrow{\nabla_{(a}\ell_{b)}}\stackrel{\Delta}{=}0$, so that $\ell^{a}$ is a Killing vector field on the $\Delta$., although the immediate neighbourhood of the hypersurface may be completely non- stationary.\\

Given this general description of a twist-free, 
shear- free and expansion- free null 
surface $\Delta$, let us now define 
the notion of a weak isolated horizon (WIH). 
The null hypersurface $\Delta$ in $\mathcal{M}$ is 
called a WIH if the following conditions hold:
\begin{enumerate}
    \item $\Delta$ is topologically $\mathbb{S}\times \mathbb{R}$,
    where $\mathbb{S}$ is a compact and connected $(N-2)$ surface. 
    \item The expansion scalar of the null normal $\ell^{a}$, given by $\theta_{(\ell)}=q^{ab}\, \nabla_{a}\ell_{b}$ is vanishing on $\Delta$, that is, $\theta_{(\ell)}\=0$.
    \item $-T_{a}{}^{b}\,\ell^{a}$ is future directed and causal, and all field equations of matter and gravitational fields hold on $\Delta$.
    \item The quantity $[\lie_{\ell}, \, \nabla_{\ula{a}}]\ell^{a}{}\=0$ acting on all vector fields tangential to $\Delta$.
    \item  The scalar field $\phi$ is Lie dragged along $\Delta$, $\lie_{\ell}\, \phi\=0$. 
\end{enumerate}
Let us provide a brief discussion on these conditions and how they might be helpful in modeling a black hole horizon. The first condition is on the topology of 
horizon cross- section. As is well known, Hawking's theorem restricts the topology of black hole horizons (for asymptotically flat spacetimes) in $4$-dimensions to $S^{2}$, although it may admit complicated topologies in higher dimensions \cite{Hawking:1971vc,Galloway:2005mf}. 
The topological requirement on $\Delta$ is stated to represent general situations. 
The condition that $\Delta$ is expansion free indicates 
that no matter or gravitational field crosses the horizon, and this is the defining requirement 
representing a non- expanding isolated null surface through which no energy momentum passes through. 
Note that not all null
hypersurface represents a WIH, expansion- freeness is crucial, the Minkowski null cone for example, is not a WIH. The third condition demands that the field equations hold on the WIH, and that
the energy- momentum tensor satisfy  
the dominant energy condition), with
$-T_{a}{}^{b}\ell^{a}$ being a future directed and null vector field. This may then be connected to the Ricci tensor through the Einstein equations. However, in higher curvature theories, 
the field equations do not connect 
the $R_{ab}$ to $T_{ab}$ and therefore, one may 
directly impose such a restriction on $R_{ab}$. 
Given these three conditions, the $\Delta$ 
only represents a non- expanding horizon (NEH). NEH cannot satisfy the laws of black hole mechanics, and one needs to restrict the geometry further to extract meaningful physics. The fourth condition restricts the connection components on the normal bundle. More precisely, if $\omega_{a}$ is the connection corresponding to the vector field $\ell^{a}$ eqn. \eqref{gradleqn1}, this
condition requires that it be Lie dragged along $\Delta$,
$\lie_{\ell}\, \omega_{a}\=0$. Since we shall be working with non- minimally coupled scalar field ($\phi$), we shall impose the restriction that the field is essentially constant along the horizon generator. 
The last condition becomes essential for a proof of 
the first law.
\\

Now that we have a horizon $\Delta$, let us begin our proof of the laws of thermodynamics. The first is the \emph{Zeroth Law:} The curvature of the connection component (also called the rotation one form) $\omega$ will play an important role. Note that the rotation one-form $\omega_{a}$ in eqn. \eqref{gradleqn1} may be written as:
\begin{equation}\label{omega_expansion}
        \omega_{a}\stackrel{\Delta}{=}-\kappa_{(\ell)} n_a + \Tilde{\omega}_{a},
\end{equation}
where $\Tilde{\omega}_{a}$ is pullback of the one form on the cross-sections, and if the horizon is non-rotating, then $\Tilde{\omega}_{a}=0$. Also, from 
$\nabla_{[a}\nabla_{b]}\ell^{c}= R_{ab}{}^{c}{}_{d}\,\ell^{d}$, one gets that 
\begin{equation}\label{Riemann tensor condition}
   \tensor{R}{_{\underleftarrow{ab}}^c_d}\,\ell^{d}\stackrel{\Delta}{=}(d{\omega})_{\underleftarrow{ab}}\,\ell^{c}\, .
\end{equation}
From eqn. \eqref{Riemann tensor condition} it is easy to show that $d\omega$ is a purely spatial form and hence it can be written as:
\begin{equation}\label{domega}
    d\omega\=C_{ij}\, m^{(i)}\wedge m^{(j)}
\end{equation}
where the scalars $C_{ij}$ are related to the Weyl tensor \cite{Chatterjee:2025jur}. From the fourth boundary  condition   $\lie_{\ell}\, \omega\=0$, it follows that:
\begin{equation}\label{wih}
  d\,\kappa_{(\ell)}\=0,
\end{equation}
where we have used eqn.\eqref{omega_expansion}, and that $\kappa_{(\ell)}$ is the acceleration of the vector field $\ell^{a}$ on the horizon. This proves the zeroth law: \emph{surface gravity is a constant on the horizon $\Delta$}.  We see that zeroth law emerges directly from the geometrical structure of the horizon and the requirement that connection component on the normal bundle is \emph{time} independent, 
or in other words, is Lie dragged along the horizon generator.\\

Let us now obtain the first law, which requires the knowledge of 
the dynamics of gravitational fields. The Lagrangian $N$-form for non minimally coupled scalar field  on a $N$ dimensional manifold $M$ is:
\begin{equation}\label{PalatiniLagrangian}
   16 \pi G \, L= f(\phi)\Sigma_{IJ}\wedge F^{IJ  }- \frac{1}{2} K(\phi) (d\phi \wedge ^*d \phi)\, -V(\phi)~ {}^{N-2} \epsilon-d[f(\phi)\,\Sigma_{IJ}\wedge A^{AJ}]
\end{equation}
where the quantity $\Sigma_{IJ}$ is constructed out of tetrads $e^{I}$ whereas $F_{IJ}=dA_{IJ}+A_{IK}\wedge A^{K}{}_{J}$ is the field strength corresponding to the connection $A_{IJ}$. The quantity $K(\phi)$ is a function of the scalar field which may be chosen to match the action to the second order metric based action \cite{carroll2019spacetime}.
The action also contains a boundary term which is similar to the Gibbons- Hawking boundary action.
\begin{equation}
    \Sigma_{IJ}=\frac{1}{(N-2)!}\,\epsilon_{IJI_1 I_2...I_{N-2}}\,\, e^{I_1}\wedge e^{I_2} \wedge...\wedge e^{I_{N-2}}
\end{equation}
The variation with respect to the connection $A_{IJ}$ gives us the following equation of motion:
\begin{equation}\label{connection eom}
    D \Tilde{\Sigma}_{I J}=(-1)^{N+1}\,d \Tilde{\Sigma}_{IJ} +\Tilde{\Sigma}_{IK}\wedge {A^K}_J + \Tilde{\Sigma}_{KJ}\wedge {A^K}_I=0
\end{equation}
 where $D$ is the gauge covariant derivative operator, $D\lambda^{I}=\partial\lambda^{I}+A^{I}{}_{J}\, \lambda^{J}$, for any internal vector $\lambda^{I}$. Also,
\begin{equation}\label{Sigma 1}
    \Tilde{\Sigma}_{IJ}:= f(\phi) \, \Sigma_{IJ}
\end{equation}
The variation with respect to the tetrads give us the Einstein equations up to a boundary term. 
Let us make the following rescaling of the co-tetrad fields: $e^I\xrightarrow{} \widehat{e}^{\,I}=p^{-1} e^I$, such that $p=1/f(\phi)^{\frac{1}{N-2}}$. We find that $A$ is the unique Lorentz connection compatible with $\widehat{e}^{\,I}$. Further, when the equation of motion \eqref{connection eom} is satisfied, the first order action is equivalent to the standard non-minimally coupled Einstein-Hilbert action up to a surface term.
The on-shell variation of the action leads to the following integral:
\begin{equation}\label{delta S}
    \delta S= \int_{\partial M= \Delta~\cup~ i^o}(-1)^{N} (  \delta \Tilde{\Sigma}_{I J}\wedge  A^{IJ})-K(\phi)~ ^*d\phi\, \delta\phi\, ,
\end{equation}
 which has to vanish for the action principle to be well- defined. The quantities at the asymptotic boundary vanish by appropriate fall- off conditions on the fields at $i^{0}$, whereas those on the horizon need to be properly evaluated. We show that this integral vanishes on $\Delta$. To show this, let us first calculate the expression of $A^{IJ}$:
\begin{equation}
\nabla_{\underleftarrow{b}}\ell^{a}=\nabla_{\underleftarrow{b}}(e^{a}_{I}\ell^{I})=(\nabla_{\underleftarrow{b}}e^{a}_{I})\,\ell^{I}+e^{a}_{I}\,\nabla_{\underleftarrow{b}}\,\ell^{I}=(\nabla_{\underleftarrow{b}}e^{a}_{I})\,\ell^{I}+e^{a}_{I} \,\tensor{A}{_{\underleftarrow{b}}}^{I}{}_{J}
\end{equation}
But the $\nabla_a$ is compatible with $\widehat{e}^{\,a}_{I}$,  i.e. $\nabla_{b}\, \widehat{e}^{\,a}_{I}=0 $, which is true on the horizon. Also, since $\widehat{e}^{\,a}_{I}=p\, e^{a}_{I}$, we get that $\nabla_{\underleftarrow{b}}e^a_I=-(\nabla_{\underleftarrow{b}}\text{ln}~p)e^a_I$. One may then rewrite:
\begin{equation}
    \omega_{b} \,\ell^{a}\=\nabla_{\underleftarrow{b}}\ell^a=-(\nabla_{\underleftarrow{b}}\ln{p})\, e^{a}_{I} \,\ell^{I} + e^{a}_{I} \tensor{A}{_{\underleftarrow{b}}^I_J}
\end{equation}
and therefore,
\begin{equation}
(\omega_b+\nabla_{\underleftarrow{b}}\ln{p})\=\tensor{A}{_{\underleftarrow{b}}^I_J}\ell^{I} n^{J}\, .
\end{equation}
So that we can write:
\begin{equation}\label{coupled connection}   \underleftarrow{A}_{IJ}\=-2(\omega+d~\ln{p})\ell_{[I} n_{J]}+C_{IJ}\, ,
\end{equation}
such that $C_{IJ}\ell^I \=0$. On the other hand,
the expression for  ${\Sigma}_{IJ}$ on horizon in terms of internal vectors can be written as:
\begin{equation}\label{Sigma}
\underleftarrow{{\Sigma}}_{IJ}\stackrel{\Delta}{=}2\,^{N-2}\epsilon\, \ell_{[I} n_{J]}
+2\alpha_k \ell_{[I} m_{J]}^{(k)}
\end{equation}
where $\alpha_{(k)}$ is a $(N-2)$ form. Using the expressions of $A_{IJ}$ from eqn. \eqref{coupled connection},  $\Sigma_{IJ}$ from \eqref{Sigma}, we get
\begin{equation}\label{Integrand}
\delta \underleftarrow{\widetilde{\Sigma}}_{IJ} \wedge \underleftarrow{A}^{IJ}
\stackrel{\Delta}{=}2 \delta(f(\phi){}^{N-2}\epsilon)\wedge(\omega+d~\ln{p})\, .
\end{equation}
So that we can write \eqref{delta S} as follows:
\begin{equation}
    \delta S=\int_{\Delta} [2 \delta\{f(\phi){}^{N-2}\epsilon\}\wedge(\omega+d~ \ln{p})]-K(\phi)~ ^*d\phi \,\delta \phi\, .
\end{equation}
  To show the above integral vanishes on $\Delta$, note that on some initial cross section $S_-$ of $\Delta$, $\delta \phi=0$ and $\delta (^{N-2}\epsilon)=0$. Since both $\phi$ and $^{N-2}\epsilon$ are Lie dragged along the horizon generator $\ell$ and also $\delta \ell\=c \ell$, $\delta \phi\=0$ and $\delta (^{N-2}\epsilon)\=0$. This proves that the surface term is zero due to the boundary conditions, on the horizon, and the action principle is well defined, hence equations of motion follow from $\delta S=0$.\\

 To obtain the symplectic structure, we proceed as follows. A variation of the Lagrangian $n$- form (in $n$- dimensions) gives the equation of motion and a boundary term: $\delta L=(\text{EOM})\, \delta (\text{fields}) +d \theta(\delta)$, with $\theta(\delta)$ being the symplectic potential, a spacetime $(N-1)$ form, but a $1$- form on phase space. The symplectic current is a closed $2$- form on phase space, obtained from the antisymmetric combination of symplectic potential \cite{Lee:1990nz, Ashtekar:1990gc}: $J(\delta_{1}, \delta_{2})=\delta_{1}\, \theta(\delta_{2})-\delta_{2}\, \theta(\delta_{1})$. From the Lagrangian in \eqref{PalatiniLagrangian}, the symplectic potential is   $\theta(\delta)=\delta \Tilde{\Sigma}_{IJ}\wedge A^{IJ}$. The symplectic current can be written as:
 \begin{equation}\label{symplectic_current}
16 \pi G \,J(\delta_{1}, \delta_{2})=(-1)^N (\delta_2 \Tilde{\Sigma}_{IJ} \wedge \delta_1 A^{IJ} - \delta_1\Tilde{\Sigma}_{IJ} \wedge \delta_2 A^{IJ})+K(\phi)(\delta_{2} {}^*d\phi\,\delta_{1} \phi-\delta_{1}\,{}^*d\phi\,\delta_{2} \phi).
\end{equation}
Let the manifold $\mathcal{M}$ be such that it is bounded by the horizon $\Delta$, two partial Cauchy slices $M_{+}$ and $M_{-}$, and the spatial infinity $i^{0}$, (see  fig \ref{fig1}). Since $J(\delta_{1}, \delta_{2})$ is closed, on $\mathcal{M}=\Delta\cup M_{+}\cup M_{-}\cup i^{0}$, we get:
\begin{eqnarray}\label{symp}
\left[\, \int_{M+} -\int_{M_{-}} -\int_{\Delta} \,\right] \, J(\delta_{1}, \delta_{2})= 0,
\end{eqnarray}
where the quantity on $i^{0}$ is taken to vanish by choice of appropriate fall- off conditions on the fields. On $M_{+}$ or $M_{-}$, the value of $J(\delta_{1}, \delta_{2})$ is as given in \eqref{symplectic_current}, but on $\Delta$, we can show that that $J$ is an exact form. To do so, let us introduce a scalar potential $\psi$ for surface gravity $\kappa$:
\begin{equation}
    \lie_\ell \psi=\kappa_{\ell}~~~\text{such that,} ~~~\psi|_{S_{-}}=0
\end{equation}
Note that $\psi$ is not unique as we can define $\psi^\prime=\psi+\xi$, such that $\lie_\ell \xi=0$, hence $\lie_{\ell} \psi^\prime=\kappa_{\ell}$. We choose $\xi=-\text{ln}\, p$ and consequently may show that:
\begin{figure}\label{manifold}
    \centering   \includegraphics[width=0.48\linewidth]{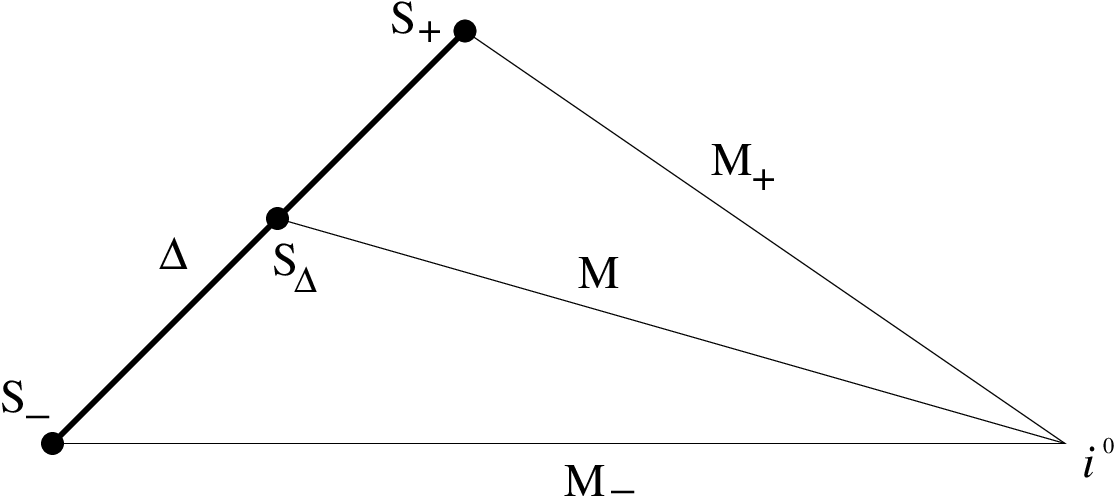}
    \caption{$M_{\pm}$ are two partial Cauchy surfaces enclosing a region of space-time and intersecting $\Delta$ in the 2-spheres $S_{\pm}$ respectively, and extend to spatial infinity $i^0$. Another Cauchy slice M is drawn which intersects $\Delta$ in $S_{\Delta}$}
    \label{fig1}
\end{figure}
\begin{equation}\label{J delta}
(-1)^N 16 \pi G\, J(\delta_1, \delta_2)|_{\Delta}
\stackrel{\Delta}{=} 
2d[\delta_{2}\{f(\phi)~^{N-2}\epsilon\}\delta_{1}\psi-\delta_{1}\{f(\phi)~^{N-2}\epsilon\}\delta_2\psi]
\end{equation}
The above result shows that the symplectic current reduces to an exact form on the horizon such that $J(\delta_1,\delta_2)|_{\Delta}=dj(\delta_1,\delta_2)$. It follows from \eqref{symp} the conserved symplectic structure can be written as:
\begin{equation}
    16 \pi G \, \Omega(\delta_1,\delta_2)=\int_M J(\delta_1,\delta_2)-\int_{S_{\Delta}} j(\delta_1,\delta_2)
\end{equation}
Therefore, the symplectic structure on the phase space of $N$- dimensional Palatini action with a non- minimally coupled scalar field admitting a WIH as an inner boundary is:
\begin{eqnarray}\label{symplectic structure}
16 \pi G\, \Omega(\delta_1, \delta_2)
&=&\int_M (-1)^N\, (\delta_2 \Tilde{\Sigma}_{IJ} \wedge \delta_1 A^{IJ} - \delta_1\Tilde{\Sigma}_{IJ} \wedge \delta_2 A^{IJ})+\int_M K(\phi)(\delta_{2}{}^*d\phi\,\delta_{1} \phi-\delta_{1}\,{}^*d\phi\,\delta_2 \phi)\\ \nonumber
&-&2\oint_{S_{\Delta}} \delta_{2}\,[f(\phi)~^{N-2}\epsilon] \delta_1 \psi-\delta_{1}[f(\phi)~^{N-2}\epsilon] \delta_2 \psi \,\, .
\end{eqnarray}
Note that this symplectic structure is independent of the hypersurface 
foliating the spacetime. \\

To determine the first law for an axisymmetric horizon, we introduce the vector field on the spacetime  $t^{a}=B_{(\ell,t)}\ell^a-\Omega_{t} \,\varrho^{a}$, where $\varrho^{a}$ is the angular Killing vector field corresponding to rotational symmetry of the cross- section. More precisely we assume that the following conditions hold true: $\lie_{\varrho}\, \omega\= 0$, $\lie_{\varrho}\, q_{ab}\=0$, $\lie_{\varrho}\, \phi\=0$, and $\lie_{\ell}\,\varrho \=0$. Let $\Omega(\delta, \delta_{t})= X_t(\delta)$, and then using the equations of motion, after a few lines of calculations, one gets :
\begin{equation}
    X_{t}(\delta)= \frac{1}{8 \pi G}\,\left[\oint_{S_{\Delta}}- \oint_{S_{\infty}}\right] \, \delta[ {}^{N-2} \epsilon ~f(\phi)]\,\kappa_{t} - \Omega_{t}\, \delta[\varrho\cdot \omega f(\phi)\,^{N-2}\epsilon ]+\delta E^{t}_{\infty}.
\end{equation}
Here, $E^{t}_{\infty}$ is the contribution to the ADM energy. In order to show that $\delta_{t} $ is Hamiltonian, we must show that $X_{t}(\delta)=\delta H_{t}$ is a closed form. This requires that the surface gravity $\kappa_t$ and angular velocity $\Omega_{t}$ must be function of area and angular momentum, respectively. Naturally, this may be written in the following form:
Here, $E^{t}_{\infty}$ is the contribution to the ADM energy.

\begin{equation}\label{first_law}
    \delta E^{t}_{\Delta}=\frac{1}{8\pi G}\,\kappa_{t}\, \delta\Tilde{a}_{\Delta}+\Omega_{t}\, \delta J_{\Delta},
\end{equation}
where $\kappa_{(t)}$ is the surface gravity associated with $t^a$. The other quantities are defined as follows: $H^{t}=E^{t}_{\infty}-E^{t}_{\Delta}$, while the quantity behaving like the area $\tilde{a}_{\Delta}$  and the angular momentum $J_{\Delta}$ is given by:
\begin{equation} \Tilde{a}_{\Delta}=\oint_{S_{\Delta}} f(\phi) ^{N-2}\epsilon
\end{equation}
\begin{equation}\label{angular momentum}
    J_{\Delta}=-\frac{1}{8\pi G}\oint_{S_{\Delta}}(\varrho\cdot \omega)\, [f(\phi)\, {}^{N-2} \epsilon\,]
\end{equation}
The requirement that $X_{t}(\delta)$ be Hamiltonian further implies that the following condition holds true:
\begin{equation}
    \frac{\partial \kappa_t(\Tilde{a}_{\Delta},J_{\Delta})}{\partial J_{\Delta}}= 8\pi G\, \frac{\partial \Omega_t(\Tilde{a}_{\Delta},J_{\Delta})}{\partial \Tilde{a_{\Delta}}}.
\end{equation}
Such a constraint ensures that the one form $X_t(\delta)$ is closed, and that $\Omega_{\Delta}$ and $\kappa_{t}$ are functions of the angular momentum $J_{\Delta}$ and the horizon area $\Tilde{a}_{\Delta}$.\\

\subsection*{Alternate action for \texorpdfstring{$4$- dimensions}{4- dimensions} and covariant phase space}

Let us now restrict our attention to $4$- dimensional
spacetimes. 
The Palatini action in eqn.\eqref{PalatiniLagrangian} with $N=4$ is not the only Lagrangian in $4$-dimensional spacetimes which describes a theory with a scalar field coupled non- minimally to gravity.
A Lagrangian similar to Holst's modification
of the GR action \cite{Holst:1995pc} may also be considered  \cite{Chatterjee:2009vd}. Note that the Holst action is 
the starting point 
for LQG, and although we shall not be utilising the quantum framework of LQG, the action has important ramifications for classical spacetimes with boundaries, as we shall discuss below.  
The non- minimal scalar coupled Holst action is given by:
\begin{eqnarray}\label{lagrangian2}
-16\pi G\gamma~L &=& \gamma\, f(\phi)\,\Sigma_{IJ}\wedge F^{IJ}~-~ f(\phi)\,e_{I}\wedge e_{J}\wedge F^{IJ}~-~\gamma~d\{f(\phi)\,\Sigma_{IJ}\wedge A^{IJ}\} \nonumber\\
&& +~d\{ f(\phi)\,e_{I}\wedge e_{J}\wedge A^{IJ}\}- 8\pi GK(\phi)^{*}d\phi\wedge d\phi +16\pi GV(\phi)\epsilon,
\end{eqnarray}
where $V(\phi)$ is a potential for the scalar field, $\epsilon$ is the $4$-dimensional volume element, and the quantity $K(\phi)$ is a scalar function 
given by:
\begin{equation}
K(\phi)=\left[1+(3/16\pi G)\left(f^{\prime 2}(\phi)/f(\phi)\right)\right].
\end{equation}
But why is it worthwhile to consider 
this $\gamma$-dependent modification
of the gravitational action? To begin, 
let us discuss some notable features 
of this action. First, 
this Lagrangian is well 
defined for $4$-dimensional spacetimes only, which follows due to 
the peculiar contraction labels of 
the Lorentz indices appearing in the tetrads ($e^{I}$) and the field strength ($F_{IJ}$). Second,
note that, as compared to the Palatini action in eqn. \eqref{PalatiniLagrangian} the theory contains a term dependent on the parameter $\gamma$, called the Barbero- Immirzi parameter. The addition of this $\gamma$-dependent action does not alter 
the classical equations of motion since they vanish due to the Bianchi identities \cite{Ashtekar:2004eh, Chatterjee:2009vd}. Therefore, the classical phase space or the symplectic structure obtained from
the Palatini action of eqn. \eqref{PalatiniLagrangian} is equivalent to 
the one obtained due to the Holst action of eqn. \eqref{lagrangian2} 
(However, in the quantum theory,
the canonical transformations generated due to these $\gamma$ dependent quantity, donot act unitarily and therefore, 
the quantum theory 
admits inequivalent 
$\gamma$- sectors \cite{Ashtekar:2004eh}). 
So, although the action  eqn. \eqref{lagrangian2} is equivalent to 
the Palatini action of eqn. \eqref{PalatiniLagrangian}, and the first law of black hole mechanics does not change,
the presence of an inner boundary admitting residual local Lorentz symmetries alters the situation (to be discussed in the next section): the $\gamma$ dependent action does give rise to additional degrees of freedom and surface charges which has important ramifications for the quantum structure of horizons.  \\

One may proceed as in the previous subsection to obtain the phase space from the action eqn. \eqref{lagrangian2}. For the $4$- dimensional spacetimes, one may use the standard Newman- Penrose null tetrads $(\ell^{a}, n^{a}, m^{a}, \bar{m}^{a})$, such that $\ell\cdot n=-1$, and $m\cdot\bar{m}=1$, with all other products
being zero. In this basis, the spacetime metric is 
given by $g_{ab}=-2\l_{(a}n_{b)}+ 2 m_{(a} \bar m_{b)}$.
The horizon $\Delta$ is a $3$- dimensional hypersurface generated by the null vector field $\ell^{a}$. $\Delta$ is foliated by $2$-spheres with $(m,\bar{m})$ being tangential to these cross-sections, and 
the area $2$-form on 
$S^{2}$ is given by ${}^{2}\epsilon=im\wedge\bar{m}$. 
One may determine the tetrads and spin- connections 
for this spacetimes in a manner similar to our discussions on $n$- dimensions:

\begin{eqnarray}
e^I_{\underleftarrow{a}}
&\=& -n_a \ell^I + m_a \bar{m}^I + \bar{m}_a m^I \\
e^I_{\underleftarrow{a}}\wedge e^J_{\underleftarrow{b}}&\=& -2(n\wedge m)_{ab}\ell^{[I}\bar{m}^{J]}-2(n\wedge \bar{m})_{ab}\ell^{[I}{m}^{J]}+2i m^{[I}\bar{m}^{J]} ~^2 \epsilon_{ab}\\
\underleftarrow{\Sigma}^{IJ} 
&\=&
2\,\ell^{[I} n^{J]} \,^2\epsilon
+ 2\, n \wedge \left(
im \ell^{[I} \bar m^{J]}
- i \bar{m} \ell^{[I} m^{J]}
\right)\\
\underleftarrow{A}_{IJ}&\=&-2(\omega+d~\text{ln}~p)\ell_{[I} n_{J]}+C_{IJ} .
\end{eqnarray}
The symplectic structure corresponding to the Holst action of 
eqn. \eqref{lagrangian2} may be obtained following
the steps leading to eqn. \eqref{Palatini_2} (see \cite{Chatterjee:2009vd} for a detail derivation):
\begin{eqnarray}\label{Palatini_2}
\Omega(\delta_{1}, \delta_{2}
)&=&\frac{1}{8\pi G\gamma}\int_{M}\left[ \delta_{1}\{ f(\phi)\,e^{I}\wedge
e^{J}\}~\wedge\delta_{2}A^{(H)}_{IJ} -\delta_{2}\{f(\phi)\,e^{I}\wedge
e^{J}\}~\wedge\delta_{1}A^{(H)}_{IJ} \right] \nn
&&~~~~~+\frac{1}{8\pi G\gamma}\int_{S_{\Delta}}\left[
\delta_{1}\{ f(\phi){}^2\mbf{\epsilon}\}~\delta_{2}\{\mu_{(m)}+ \gamma\psi_{(\l)}\} -
\delta_{2}\{ f(\phi){}^2\mbf{\epsilon}\}~\delta_{1}\{\mu_{(m)} + \gamma\psi_{(\l)}\}\right]\nonumber\\
&&~~~~~~~~~~~~~~~~~~~~~~+\int_{M}\,K(\phi)\left[\delta_{1}({}^{*}d\phi)\delta_{2}\phi~-~\delta_{2}({}^{*}d\phi)\delta_{1}\phi\right],
\end{eqnarray}
where $A^{(H)}_{IJ}=(1/2)[A_{IJ}-(1/2\gamma)\,\epsilon_{IJ}{^{KL}\, A_{KL}}\, ]$. This phase space does give rise to the standard laws of black hole mechanics, as has been already shown in \cite{Chatterjee:2009vd}, and shall not be repeated here. We shall use this symplectic structure for our calculations for the quantum theory in the next section.\\

\section{The local Lorentz symmetries on WIH and Hamiltonian charges}

The Lagrangian of eqn. is a first order theory based on the tetrad- connection formalism.
This theory is equivalent to the second order metric theory if one imposes the zero torsion condition. However, the first order theory is more natural since gravity is known to be invariant under 
diffeomorphisms and well as local Lorentz transformations. The first order 
theory of eqn. \eqref{lagrangian2} cleanly captures 
the action of both the diffeomorphisms as well as local Lorentz group of 
transformations $\text{SL}(2,\mathbb{C})$. The co-tetrads and the connection transform 
under a Lorentz transformation in the following way.
\begin{eqnarray}
e^{I}&\rightarrow &\Lambda^{I}{}_{J}\,e^{J}\\
A^{IJ}&\rightarrow& (\Lambda^{-1})^{I}{}_{K}\,A^{KL}\,\Lambda_{L}{}^{J}+ (\Lambda^{-1})^{I}{}_{K}\,d\Lambda^{KJ}
\end{eqnarray}
where ${\Lambda^I}_{J}$ is the Lorentz transformation matrix. 
When boundaries are present, 
the full symmetry 
group is broken
and only the residual transformations must be considered, which preserve the boundary conditions. This leads to
the emergence of new physical degrees of freedom (called edge states) associated specifically to 
these boundaries. For example, mass and angular momentum
may be considered as global charges arising out of broken diffeomorphisms at the boundaries, like 
the asymptotic infinities. Similarly, 
the presence 
of a WIH (considered as an inner boundary of 
the spacetime) also breaks $\text{SL}(2,\mathbb{C})$ to $\text{ISO}(2)\ltimes \mathbb{R}$ \cite{Basu:2010hv,Chatterjee:2020iuf}. The question is the following: what are residual symmetries which map $\Delta\mapsto \Delta$? What are the charges which emerge from these broken gauge symmetries?\\

Let us answer the first question. 
The set of all Lorentz transformations belonging to the group $\text{SL}(2\mathbb{C})$ consists of 
the following set of six transformations: (a)
a boost in the null $(\ell-n)$ plane, 
(b) rotation in the null plane of $(m-\bar{m})$ 
(c) a boost and a rotation
in the $(\ell-m)$ plane with $n$ fixed and
(d) a boost and a rotation in
the $(n-m)$ plane (keeping $\ell$ fixed). 
One may show \cite{Basu:2010hv,Chatterjee:2020iuf,Chatterjee:2015lwa}, that only the first three transformations are useful and map a WIH to a WIH. 
The Lorentz matrices associated with these transformations are respectively
\begin{align} \Lambda_{IJ}=&-\xi\l_In_J-\xi^{-1}n_I\l_J+2m_{(I}\bar
m_{J)},\label{L1}\\
\Lambda_{IJ}=&-2\l_{(I}n_{J)}+(e^{i\theta}m_I\bar m_J+c.c.),\label{L2}\\
\Lambda_{IJ}=&-\l_In_J-(n_I-cm_I-\bar cm_I+|c|^2\l_I)\l_J +(m_I-\bar c\l_I)\bar m_J+(\bar m_I-c\l_I)m_J .\label{L3}\end{align}
The generators corresponding to these transformations are given by the following quantities:
\begin{eqnarray} 
&B_{IJ}=(\partial\Lambda_{IJ}/\partial\xi)_{\xi=1}=-2\l_{[I}n_{J]},
\label{lbb}\\
&R_{IJ}=(\partial\Lambda_{IJ}/\partial\theta)_{\theta=0}=2im_{[I}\bar
m_{J]},\label{lbr}\\
&P_{IJ}=(\partial\Lambda_{IJ}/\partial{\rm Re}\,c)_{c=0}=2m_{[I}\l_{J]}+2\bar
m_{[I}\l_{J]},\label{lbp}\\
&Q_{IJ}=(\partial\Lambda_{IJ}/\partial{\rm Im}\,c)_{c=0}=2im_{[I}\l_{J]}-2i\bar
m_{[I}\l_{J]},\label{lbq}
\end{eqnarray}
where $B,R$ generate the transformations (a) and (b) respectively, while $P,Q$ generate (c). 
A straightforward calculation gives their Lie brackets
\begin{align} &[R,B]=0,\quad [R,P]=Q,\quad [R,Q]=-P,\nonumber\notag\\
&[B,P]=P,\quad [B,Q]=Q,\quad [P,Q]=0,\label{iso2}\end{align}
where 
the symbol $[M,N]_{IJ}=M_{IK}\,N^K{}_J-N_{IK}\, M^K{}_J$. 
This is the Lie
algebra of $\text{ISO}(2)\ltimes\mb R$ where the symbol $\ltimes$ stands for the semidirect
product. Note that
$\text{ISO}(2)$ is the little group of the Lorentz group which maps a null vector to itself and therefore, it is 
not surprising that the boundary conditions on $\Delta$ are invariant only under this subgroup of
local Lorentz group. So to summarise,
$R$ generates Euclidean rotations
in the $(m-\bar{m})$ plane, $P$ generates rotation in the $(\ell-{m})$,
$Q$ generates rotation in the $(\ell-\bar{m})$; while, $B$ generates $\mb R$, scaling transformations of  $(\ell-{n})$. \\

To answer the second question, we would like to construct the Hamiltonian charges for 
these transformations. 
We shall show next 
that the charges corresponding to $B$ and $R$
which should generate boost and angular momentum on 
the phase- space respectively, are related to horizon area. To obtain this, let us consider the variations of 
the co-tetrads and the connection due to infinitesimal Lorentz transformations, $\Lambda^{I}{}_{J}=(\delta^{I}{}_{J}+\lambda^{I}{}_{J})$. These
are given by: 
\begin{eqnarray}
\delta_\lambda e^I&=&\lambda^{I}{}_{J}\,e^{J}\label{delta_e}\\
\delta_\lambda A^{IJ}&=&-d\lambda^{IJ} -A^{IK}\lambda_{K}{}^{J}-A^{JK}\,\lambda^{I}{}_{K}\label{delta_a},
\end{eqnarray}
where $\epsilon_{IJ}$ are generators of 
\eqref{lbb}-\eqref{lbq}.
We also require the expression for the variation of $\Sigma_{IJ}$ and that of $(e_{I}\wedge e_{J})$. 
After a bit of algebra, one can show that \cite{Chatterjee:2020iuf},
\begin{eqnarray}\label{lortrans2}
\delta_{\lambda} \Sigma_{IJ}
&=&\lambda_{KI}\,\, \Sigma_{J}{}^{K}-\lambda_{KJ}\,\,\Sigma_{I}{}^{K},\label{delta_sigma}\\
\delta_\lambda(e_{I}\wedge e_{J})&=&\lambda{}_{IK}\,\,e^{K}\wedge e_{J}+\lambda{}_{JK}\,\, e_{I}\wedge e^{K}\label{delta_ee1}.
\end{eqnarray}

Using these transformations for the tetrads and the connection variables under Lorentz transformations,
we determine the Hamiltonian charges as follows. We use the bulk part of the symplectic structure of 
equation \eqref{Palatini_2}, keeping in mind that
the scalar field is a constant on the horizon $\Delta$.
The contribution from the surface symplectic structure are sometimes important for the proof of first law of black hole mechanics and is crucial 
\cite{Chatterjee:2008if}, although for Lorentz transformations, all contributions from the boundary symplectic structure vanish. So the bulk part of 
symplectic structure of eqn. \eqref{Palatini_2}, 
along with eqns. \eqref{lortrans2} and 
\eqref{delta_ee1} gives us:
\begin{eqnarray} \label{symp1}
\Omega_{B}(\delta_{\lambda},\delta)=- \frac{1}{16\pi G\gamma }\int_{S_{\Delta}}\delta\,\left[f(\phi_{0})\,\,e_{I}\wedge e_{J}-\gamma f(\phi_{0})\,\Sigma_{IJ}\,\right]\,\ \lambda^{IJ}.
\end{eqnarray}
Note that here, we have fixed the value of the scalar field $\phi=\phi_{\Delta}$ on the horizon. 
For $\lambda_{IJ}=R_{IJ}=2im_{[I}\bar m_{J]}$, the only contribution comes
through the $\gamma$ dependent symplectic structure, and one obtains the Hamiltonian charge for local Lorentz rotations:
\begin{eqnarray}\label{ham_charge_rot}
\Omega_{B}(\delta_{R},\delta)=- \frac{1}{8\pi G\gamma }\int_{S_{\Delta}}\delta\, {}^{2}\epsilon
=- \delta\left[\frac{f(\phi_{\Delta}\,)\mathcal{A}}{8\pi G\gamma }\right]\equiv\delta(-\mathcal{J}),
\end{eqnarray}
where, for notational simplification, we denote 
the charge by $(-\mathcal{J})$, negative of 
the angular momentum corresponding 
to local Lorentz rotations. 
So, if we define
$f(\phi_{\Delta})\, \mathcal{A}=\tilde{\mathcal{A}}$, and call it the modified area, then 
\begin{equation}\label{area_spectrum}
\mathcal{J}=\frac{\tilde{\mathcal{A}}}{8\pi G\gamma}\, ,
\end{equation}
is the generator
of local Lorentz rotations on the phase- space of isolated horizons.
For $\lambda_{IJ}=B_{IJ}=-2l_{[I}n_{J]}$, we shall denote the charge by $\mathcal{K}$ (boost) on 
the horizon and the only contribution comes
through the $\gamma$- independent symplectic structure eqn. \eqref{symp1}.
\begin{eqnarray}\label{ham_charge_boost}
\Omega_{B}(\delta_B,\delta)=\frac{1}{8\pi G }\int_{S_{\Delta}}\delta\, {}^{2}\epsilon
=\delta\left[\frac{\tilde{\mathcal{A}}}{8\pi G}\right]\equiv\delta(\mathcal{K}).
\end{eqnarray}
Again, $\mathcal{K}=(\tilde{\mathcal{A}}/8\pi G)$ is the generator of local Lorentz boosts on 
the phase- space of isolated horizons. The Hamiltonian charges for $\lambda_{IJ}=P_{IJ}$ or $\lambda_{IJ}=Q_{IJ}$, which we shall denote by $\mathcal{P}$ and $\mathcal{Q}$ respectively, vanish on the WIH phase- space and therefore,
do not generate evolutions. 
\\

From the equations, eqn. \eqref{area_spectrum} and eqn. \eqref{ham_charge_boost}, one obtains 
the \emph{linear simplicity constraint} 
$\mathcal{K}=\gamma \mathcal{J}$ which has important implications for quantum gravity, but has been argued from completely different perspective  \cite{Rovelli:2013osa}. Secondly, 
the area of the horizon,  modified by the scalar field, is linked to the angular momentum $\tilde{\mathcal{A}}=8\pi G \gamma \, \mathcal{J}$, as well as the local Lorentz boost, $\tilde{\mathcal{A}}=8\pi G\,\mathcal{K}$. These equations 
shall play a fundamental role in 
our understanding
of quantum structure of horizons. We shall show 
that the quantum states residing on the
horizon carry the finite dimensional representation of $\text{iso}(2)$, and are eigenstates of $\mathcal{J}$.
These states are labeled by integers 
and consequently, the equation $\tilde{\mathcal{A}}=8\pi G \gamma \,\mathcal{J}$ gives rise to a discrete and equidistant spectrum
of $\tilde{\mathcal{A}}$.
%




\section{Area spectrum and entropy of WIH}
To obtain the area spectrum, we proceed as follows.
First, we note that the algebra of vector fields is 
mapped to the algebra of Hamiltonian charges on 
the phase -space, and hence 
there is no central charge. In other words, the Poisson brackets of the Hamiltonian charges, obtained from
the symplectic structure \eqref{Palatini_2}, follow 
the following algebra:
\begin{equation}
    \{\mathcal{J}, \, \mathcal{K}\}=0,~~  \{\mathcal{J}, \, \mathcal{P}\}=\mathcal{Q}, ~~\{\mathcal{J}, \, \mathcal{Q}\}=\mathcal{-P}.
\end{equation}  
The corresponding quantum algebra is given by:
\begin{equation}
    [\mathcal{J}, \, \mathcal{K}]=0,~~  [\mathcal{J}, \, \mathcal{P}]=i\hbar\mathcal{Q}, ~~[\mathcal{J}, \, \mathcal{Q}]=-i\hbar\mathcal{P}.
\end{equation}  
Naturally, on the horizon cross-section, the quantum states are in the representation of $iso(2)$. 
The operator $\mathbb{P}\equiv \mathcal{P}^{2}+\mathcal{Q}^{2}$ is a Casimir of the algebra and therefore, $\mathbb{P}$ and $\mathcal{J}$ commute.
Hence, the states must be labelled by the eigenvalues of $\mathcal{P}$ 
and $\mathcal{J}$. Let the eigenvalues of $\mathbb{P}$ 
and $\mathcal{J}$ are $p$ and $j$ respectively, therefore these 
states shall be denoted by
$|p,\, j\rangle$. The operators $\mathcal{P}$
and $\mathcal{Q}$ may be combined to create ladder operators for the algebra $\mathcal{P_{\mp}}=\mathcal{P}\mp i \mathcal{Q}$, which steps up and down the eigenvalues of $\mathcal{J}$,  such that $\mathcal{P_{\mp}} |p,\, j\rangle =\mp\hbar|p,\, j\mp 1\rangle$ \cite{Chatterjee:2020iuf}. 
 We have already argued in the previous section that for
the WIH, the generators corresponding to $P_{IJ}$
and $Q_{IJ}$, denoted by $\mathcal{P}$
 and $\mathcal{Q}$ respectively, vanish. This is a reflection of the fact that by definition, 
 the horizon area of WIH 
 does not increase. Therefore, all 
 the physical states 
of the horizon must satisfy this constraint (that area 
should not increase) for which the states must have $p=0$, and must be labeled 
by $j$ only. So, the irreducible representations of the $iso(2)$ algebra useful for the WIH
are one-dimensional. The states belonging to the horizon are the eigenstates of $\mathcal{J}$ labeled by integer $j=n$, with $n\in \mathbb{N}$ \cite{Weinberg:1995mt}. To summarize, we have obtained the states
which belong exclusively on the horizon (are independent of the bulk or how the horizon is 
embedded) and carry integer labels.\\

Second, as we have shown above, on
the phase space containing WIH as inner boundary, the 
states on the horizon are eigenstates of 
the operator $\mathcal{J}$ with integer values.
This label may be used to determine the spectrum of
the area operator. Using the eqn. \eqref{ham_charge_rot}, we obtain (using $c=1$ units):
\begin{equation}\label{area_oper}
    \tilde{\mathcal{A}} \, | j \,\rangle= 8\pi G\gamma \mathcal{J}\, | j \,\rangle =  8\pi n\gamma \,\ell_{P}^{2} \, | j \,\rangle .
\end{equation}
The area eigenvalues (which we shall denote by $\mathcal{A}$ ), are then given by:
\begin{equation}
\mathcal{A}=\frac{8\pi n\gamma \,\ell_{P}^{2}}{f(\phi_{\Delta})}.
\end{equation}
This is similar to the result of \cite{Ashtekar:2000eq}, with 
the eigenvalue $[(8\pi \gamma \,\ell_{P}^{2})/f(\phi_{\Delta})]\sqrt{j(j+1)} $ where $j$ is half interger valued. Such a condition arises from quantisation requirement on the level of the Chern-Simons theory at 
the horizon boundary, and
is essential for quantising this topological theory. In the present scenario, the area spectrum arises naturally due to geometry of the WIH formalism. The spectrum of the area operator is equidistant, which is consistent with the proposal in \cite{Bekenstein:1995ju, Alekseev:2000hf} or that obtained from quasinormal modes \cite{Dreyer:2002vy}. \\

Third, we now use the area spectrum to obtain the entropy of a WIH with a fixed classical cross- sectional
area $\tilde{\mathcal{A}}_{\Delta}$.
Given the description of a
quantum state on WIH elaborated above, we shall replace the notion 
of a classical horizon and assume 
that the horizon WIH is a 
quantum state  $|J\rangle$. 
Since all states of WIH 
belong to the representation of $iso(2)$, the label of the state $J\in \mathbb{N}$.
Naturally, the classical 
area is
an eigenvalue of the area operator eqn. \eqref{area_oper}, such that:
\begin{equation}
\tilde{\mathcal{A}} \, | J \,\rangle=[8\pi \gamma \,\ell_{P}^{2}/{f(\phi_{\Delta})}]\, \mathcal{N} | J \,\rangle \, \equiv
\tilde{\mathcal{A}}_{\Delta}\, | J \,\rangle \, 
\end{equation}
where the number $\mathcal{N}$ is 
a large integer for which 
the classical area is obtained. 
We assume that the surface $S_{\Delta}$ of the WIH is tessellated by patches of area, much like the surface of a soccer ball or a map of the earth. Since the tessellated patches are at particular positions on
the $2$-sphere, they must be subjected to diffeomorphism constraints \citep{Ashtekar:2000eq}. We shall fix this constraint by coloring these tessellations, which makes these patches distinguishable. Since each patch of area corresponds to a particular state belonging to $iso(2)$, we view the black hole horizon as a composite system created out of these states.  The Hilbert space of the horizon surface is a tensor product of all these states.  The area of each such tessellated patch is labeled by integers, so that the total state $|J\rangle$ must be described by a tensor product structure $|J\rangle=\otimes_i|j_{i}\rangle$ where $i$ labels the patches. 
The area operator is taken to be acting on tessellations as follows: $\tilde{\mathcal{A}}=\oplus_{i}\tilde{\mathcal{A}}_{i}$,
so that each patch contributes an area $8\pi \gamma \ell_{p}^{2}\,n_{i}$. Note that  $\mathcal{N}=
\sum_{i} n_{i}$. 
To obtain the black hole entropy, we shall use the 
microcanonical ensemble and determine the number of independent ways the configurations $\{n_i\}$ is  obtained such that for fixed $\mathcal{N}$ 
(fixed area phase space) the equation $\sum_{i}\, n_{i}=\mathcal{N}$ holds true.  \\

Now to obtain the counting, we first note that
the horizons have a fixed area (or fixed $\mathcal{N}$). We assume that in the partition of $\mathcal{N}$, the number $n_{i}$ is 
shared among $s_{i}$ patches. 
This implies that $\sum_{i}s_{i}n_{i}=\mathcal{N}$, where $\sum_is_i$ is the total number of tessellations. 
So the total number of independent configurations is given by
\begin{equation}
\Omega(\mathcal{A})=\frac{(\sum_is_i)!}{\prod_is_i!}.
\end{equation}
To obtain the most probable configurations, 
we vary $\ln\Omega$ subject to the constraint $\delta\sum_{i}s_{i}n_{i}=0$. 
One obtains 
\begin{equation}
s_{i}=(\sum_is_i)\exp(-\lambda n_i)
\end{equation}
where $\lambda$ may be determined from the constraint $\sum_i\exp(-\lambda n_i)=1$ where $n_i=1,...,\mathcal{N}$. This sum may be easily evaluated and gives 
\begin{equation}
\lambda=\ln 2-2^{-\mathcal{N}}+o(2^{-2\mathcal{N}})
\end{equation}
for large $\mathcal{N}$. One may obtain the entropy from $\mathcal{S}=\lambda \mathcal{N}$.
%
Assuming that the black hole has large area, the entropy is given by
\begin{equation}\label{entropy}
\mathcal{S}(\mathcal{A})=\frac{\tilde{\mathcal{A}_{\Delta}}\ln 2}{8\pi\gamma\ell_{p}^{2}}.
\end{equation}
For the choice $\gamma=\ln(2)/2\pi$, the leading order Bekenstein-Hawking result is obtained. For small area, one also obtains an exponentially suppressed corrections to the classical result of $\exp[{-\tilde{{\mathcal A}_{\Delta}}\ln 2/8\pi\gamma\ell_p^2}]$.
This exponential suppression have been obtained in \cite{Chatterjee:2020iuf} and in string computations through non- perturbative corrections 
\cite{Dabholkar:2014ema}.\\

How does this calculation and the value of 
black hole entropy compare with previous 
calculations, particularly with those 
of LQG? Here, we must point out 
that a direct comparison is not easy
since there is a difference between
our approach presented here and those 
followed in the LQG literature. First, 
as pointed out in the introductory 
section [the paragraph containing eqn. \eqref{area_lqg}], the LQG method requires a Hilbert space comprising of both the bulk and the boundary Hilbert spaces, $\mathcal{H}=\mathcal{H}_{V}\otimes \mathcal{H}_{S}$. The surface Hilbert space contains states of 
the $U(1)/ SU(2)$ Chern- Simons theory on a punctured sphere, whereas the bulk Hilbert space is formed out of the spin network states. One gets the physical Hilbert space 
by imposing the $F-\Sigma$ equation on these states. On the other hand, our approach followed here completely decouples the bulk, both classically as well as quantum mechanically. More precisely, the WIH formalism allows 
us to construct 
an isolated null hypersurface mimicking 
a classical black hole horizon in equilibrium, whereas our construction of horizon quantum states too does not require any dependence on the bulk geometry, and follows quite naturally from the algebra of charges on the horizon. 
Therefore, in contrast to the LQG Hilbert space, the Hilbert space we use contains states localised on the horizon only. Second, the microcanonical calculation of LQG uses the square root area spectrum of LQG, eqn. \eqref{area_lqg}, along with the total projection constraint $\sum_{i}\, m_{i}=0$. Using these equations, the leading value of black hole entropy has been calculated following
a variety of techniques 
\cite{Ghosh_2005, Ghosh:2004rq, Ghosh:2004wq, BarberoG:2008dee,BarberoG:2022ixy, Domagala:2004jt, Mitra:2009ba}, and lead to different values for the $\gamma$- parameter. Our calculation, in contrast, is a straightforward application of the equidistant area spectrum and does not need 
additional constraint on the states. One may argue that the formulation 
does not arise
from an underlying theory of quantum gravity, and may only be considered as an effective quantum description of the 
localised horizon microstates, which explains the origin of black hole entropy. However, our results show that whatever the theory of quantum gravity may be, the quantum horizon must be formed out of discrete areas, a fundamental 
feature which follows directly from local Lorentz symmetry of the horizon.  \\

\section{Discussions}
This paper obtains a number of results in the context of a theory of gravity where scalar field is coupled non- minimally to gravitational fields. For this theory, we discuss the formulation of classical laws of horizon mechanics, and in particular the zeroth and the first laws due to the WIH formulation in $N$- dimensions. 
The zeroth law arises due of some
constraints imposed on the connection on the normal bundle, while the first law is a necessary and sufficient condition for a well defined Hamiltonian evolution on the phase- space admitting WIH as an inner boundary. Our proof of the first law for higher dimensions follows similar methodology as for the four dimensions. For $4$- dimensions, we show that the phase space of WIH admit Hamiltonian charges for local Lorentz transformations as well. 
More precisely, we show that local Lorentz rotations and boosts belonging to $\text{ISO(2)}\ltimes\mathbb{R}$ (the symmetry algebra on $\Delta$)
give rise to charges proportional to the horizon cross- sectional area modified by the scalar field (other minimally coupled fields do not contribute since they do not appear in the boundary symplectic structure). 
Using this definition of area in terms of the Lorentz generators, we have established that the spectrum of area operator is naturally discrete even for a non- minimally coupled theory. This area spectrum
is given by: $[(8\pi \gamma \,\ell_{P}^{2})/f(\phi_{\Delta})]n $ where $n$ carry integer values. The conclusion is reached without the need of any particular quantum theory of gravity like LQG or string theory. Furthermore, the eigenstates of the area operator are attached to the WIH boundary and do not depend on the bulk geometry. This is an advantage of this formulation since one may argue that the fields in the bulk do not affect the black hole entropy. Such a 
feature is useful since our formulation, unlike string theory or LQG, requires that the states on the horizon remain isolated from the bulk, both classically as well as quantum mechanically. The computation of entropy using the microcanonical fixed area ensemble reveals the standard Bekenstein- Hawking area law for large black holes as well as an exponential suppression for small Planck size 
black holes. Note the absence of log corrections. This is not obvious, but one may argue that the such corrections only arise ( at least in LQG) if one considers 
large area with large Chern- Simons level \cite{Ghosh:2012jf}. How does our computation of the area spectrum compare with other proposals? Our calculation shows that the simplified \emph{it from bit}
picture is a consistent formulation.
Also, our area spectrum is similar to the Bekenstein- Mukhanov picture of horizon microstates \cite{Bekenstein:1995ju} or the proposals based on quasinormal modes
\cite{Dreyer:2002vy} or from \cite{Alekseev:2000hf}. Alternative proposals have been studied in \cite{PhysRevD.104.084049, PhysRevD.105.044046}.\\

There is a major drawback of our 
approach, it is tied to $4$- dimensions, and an extension to higher dimensions shall be useful. At this point, let us emphasize the role of spacetime dimensions in our calculations. First, our proof of the first law of black hole mechanics has been carried out for arbitrary $N$-dimensional spacetime eqn. \eqref{first_law}. This proof uses the Palatini action for $N$-dimensional spacetime, which shows that first law holds naturally for all spacetime dimensions. Hence, one would expect that the interpretation of the black hole as a thermal object should also continue to hold for arbitrary spacetime dimensions as well. This would require that black holes have an entropy proportional to their area and admit temperatures proportional to their surface gravity, and it is here, that our formalism breaks down. We could only argue the existence of black hole entropy arising out of horizon microstates in four dimensional spacetime. More precisely, our formalism fails to give microscopic explanation of the origin of black hole entropy beyond $4$ dimensions. The reason for this should be obvious from our calculations in the previous sections: first, the Holst like action written in eqn. \eqref{lagrangian2} only exists when spacetime dimensions $N=4$. A consequence of this is that, in the presence of inner boundaries which admit residual local Lorentz invariance, one obtains horizon area 
as surface charges. Furthermore, the algebra of these charges also follow the algebra of vector fields generating these charges and follow the $iso(2)$ algebra peculiar to the representations of light like vector fields in a $4$ dimensional spacetime \cite{Weinberg:1995mt}. Naturally, the quantum states residing on the horizon are labeled by the representation of this $iso(2)$ algebra. The counting of microstates and the enumeration of the black hole entropy, which utilizes these states, is also based on 4-dimensional spacetime. So, to extend this formalism to higher dimensional spacetimes, new variables may be 
needed which are conducive to quantisation techniques used here. \\

\textbf{Acknowledgements:} AG acknowledges the financial support in the form of fellowship from Central University of Himachal Pradesh. SD acknowledges the financial support provided by DST vide Grant No. DST/INSPIRE Fellowship/2020/IF200537.




\end{document}